\documentclass[preprint,showpacs,prl]{revtex4}
\usepackage {graphicx}

\begin{document}

\title {\bf Lattice Resistance to Dislocation Motion at the Nanoscale}

\author {A. Dutta$^\dagger$, M. Bhattacharya$^\ddagger$, P. Barat$^{\ddagger
\footnote{Email:pbarat@veccal.ernet.in}}$, 
P. Mukherjee$^\ddagger$, N. Gayathri$^\ddagger$, G. C. Das$^\dagger$}

\affiliation {$^\dagger$School of Materials Science and Technology, Jadavpur University,
Kolkata 700 032, India\\
$^\ddagger$Variable Energy Cyclotron Centre, 1/AF, Bidhannagar, Kolkata 700 064, India}

\date{\today}

\begin{abstract}

In this letter we propose a model that demonstrates the effect of free surface on the lattice 
resistance experienced by a moving dislocation in nanodimensional systems. This effect manifests in
an enhanced velocity of dislocation due to the proximity of the dislocation line to the surface.
To verify this finding, molecular dynamics simulations for an edge dislocation in bcc molybdenum 
are performed and the results are found to be in agreement
with the numerical implementations of this model. The reduction in this effect at higher stresses 
and temperatures, as revealed by the simulations, confirms the role of lattice 
resistance behind the observed change in the dislocation velocity.

\end{abstract}
\maketitle

Mechanical properties of crystalline solids are extensively dependent on the 
dynamics of dislocations in them. This is true even for the nanoscale materials, however, 
the dynamics is often found to be influenced by the presence of the boundaries, a behavior that is 
strikingly different from that observed in the bulk 
\cite{steffen,schmid,nicola,tang,afanasyev,greer,gutkin,suresh}. 
Finite nanoscale systems possess large free surfaces. Thus, the surface plays 
an important role in the dynamics of dislocation in nanomaterials. The effects of finite length and the 
termination of a dislocation line at free surfaces have been reported earlier \cite{lin, bitzek}. 
In addition,
many complex phenomena e.g. interaction of dislocation with the surface acoustic waves \cite{maurel}, 
change in the effective mass of a dislocation near a free surface \cite{hirth}, and drastic rise 
in the velocity of dislocation near the free surface of a semiconductor presumably due to smaller 
doping concentration \cite{review} have been intensively studied in the context of surface effects 
on the dynamics of dislocations. However, the possible role of a free surface in changing the 
lattice resistance on a dynamic dislocation has not been taken into account to date.
A dislocation core near a free surface observes a different 
surrounding as compared to the one, which is present deep inside the crystal and therefore, 
is expected to experience a different lattice resistance in its dynamic state.
Consequently, it would respond differently under the same loading conditions. 
Such a variation in the lattice resistance would be reflected in the velocity of dislocation, 
which would eventually play a fundamental role in the mechanics of materials
at low dimensions.\

Aiming to explore the possibility of the aforementioned surface effects on moving dislocations, 
we introduce a model. Using this model we report for the first time, a significant change in 
the velocity of dislocation due to its proximity to free surfaces when it is moving with its 
line direction and velocity parallel to the free surfaces. Furthermore, molecular dynamics (MD) 
simulations have been carried out and the results are found to agree 
with those obtained from the numerical implementation of this model. In addition, this model 
exhibits the potential to separate out the lattice resistance from the overall drag experienced
by a moving dislocation.\

A moving dislocation core passes through periodically varying potential caused by the 
discreteness of the crystal lattice. The potential gradient opposing the displacement 
of the core gives rise to the lattice resistance experienced by the dislocation \cite{hirth,review}. 
Our model treats this phenomenon from a novel standpoint. Instead of focusing on the 
displacement of the moving core of a dislocation explicitly, we concentrate on the dynamic 
change in the displacement field experienced by the atoms, due to the movement 
of the dislocation. This model considers the fact that the potential of an atom in a lattice 
depends on the atomic configuration of the whole lattice. 
As a result, a change in the lattice resistance is expected if free surfaces are introduced. 
On the contrary, other primary drag mechanisms like phonon and electron drags \cite{hirth,review,alshits}
on the moving dislocations are expected to be less sensitive to the presence of the dislocation core 
in the vicinity of a free surface. Thus, in our model, the net drag force can be split into two 
components, one of which is strongly influenced by the presence of a dislocation core near a free surface 
while the other is not. The overall drag coefficient $B$ and the dislocation velocity 
$v$ are related as \cite{review},
\begin{equation}
v(t) =  \frac{\tau b}{B} (1-e^{-Bt/m^*})
\end{equation}
where $\tau$ is the applied shear load, $b$ is the magnitude of the Burgers vector, 
$t$ is the time elapsed after the dislocation starts moving and $m^*$ is the effective mass 
per unit length of the dislocation line \cite{hirth}. The expected change in the drag 
coefficient $B$ due to the introduction of the free surfaces leads to the prediction of an 
altered terminal velocity of dislocation, $v_0=\tau b/B$.

The movement of a dislocation due to the applied force per unit length of the dislocation 
line ($\tau b$), changes the displacement field felt by the atoms in the lattice. This is 
resisted by an equal and opposite drag force experienced by the dislocation when it attains 
its terminal velocity. In the model, we assume that the resistance to any change in the 
displacement field of an atom due to its interaction with other atoms of the lattice, contributes 
to the drag force due to the lattice resistance. Cumulative contributions from all the lattice 
atoms give the overall lattice resistance. Nevertheless, each atom undergoes a different change 
in the displacement field, and hence should contribute differently to this drag force. Thus, 
there is need for a contribution function that can express the role of the atoms in 
determining the net lattice resistance. In order to investigate the surface effects at the
nanoscale, a thin film is an ideal system as it provides infinitely large free surfaces 
with confinement along its thickness.

Consider a dislocation moving with the terminal velocity $v_0$ along the positive $x$ direction 
in a thin film bounded by the top and bottom surfaces in the $y$ direction. 
Fig. 1(a) illustrates the configuration of the moving dislocation with the line direction of 
the dislocation along the $z$ axis. At an arbitrary time $t$ the dislocation line passes through 
the origin O. The
position vector ${\bf r}_{ij}^d(t)$ of the ${ij}^{th}$ atom $A_{ij}$ in the lattice is given by
\begin{equation}
{\bf r}_{ij}^d (t) = {\bf r}_{ij}^c + {\bf u}_{ij} (t, {\bf r}_{ij}^c)
\end{equation}
where ${\bf r}_{ij}^c$ is the position vector of $A_{ij}$ in the perfect crystal and 
${\bf u}_{ij}(t, {\bf r}_{ij}^c)$ is
the corresponding displacement of the atom in the presence of the dislocation at time $t$. In a small
time interval $\delta t$, the dislocation line proceeds by a distance $\delta x = v_0 \delta t$ (refer
Fig. 1(b)). The net force due to the lattice resistance is given by

\begin{equation}
F_s^{(lattice)} =  \sum_{i=s_1}^{s_2}\sum_{j=-\infty}^{\infty}\phi_{ij} ,
\end{equation}

\noindent
where $\phi_{ij}$ denotes the contribution of the atom $A_{ij}$ to the lattice resistance. We assume 
a simple proportionality
relation between $\phi_{ij}$ and the change in the position vector of $A_{ij}$ in time $\delta t$ as

\begin{equation}
\phi_{ij} =  \kappa \left|{{\bf r}_{ij}^d (t+\delta t) - {\bf r}_{ij}^d (t)}\right|,
\end{equation}

\noindent
where $\kappa$ is the proportionality constant for a given loading condition. In terms of $\delta t$ 
as the unit of time, 				 
\begin{equation}
F_s^{(lattice)} =  B_s^{(lattice)} v_0,
\end{equation}
where  
\begin{equation}
B_s^{(lattice)} = \kappa\sum_{i=s_1}^{s_2}\sum_{j=-\infty}^{\infty}\left|\sum_{n=1}^{\infty}
\frac{v_0^{n-1}}{n!} \left(\frac{\partial^n {\bf u}_{ij}}{\partial (\delta x)^n}\right)_{\delta x=0}\right|.
\end{equation}	

\noindent
A different value of the drag coefficient $B_s^{(lattice)}$ is 
quite obvious following Eq. (6) due to the change in the summation limits of $i$ representing the 
thickness of the film. A reduction in the lattice resistance for a thinner film is indicative of 
an enhanced velocity of dislocation. Similar changes are expected due to the variation in position 
of the dislocation line along the film thickness. MD simulations are carried out to verify such effects. \

A typical simulation starts with a virtual freestanding thin film of single crystal bcc 
molybdenum created using the Finnis-Sinclair potential \cite{finnis}. The simulation cell is shown 
in Fig. 2 with its $x$, $y$ and $z$ axes along $<$111$>$, $<$$\bar{1}$01$>$, and $<$1$\bar{2}$1$>$ 
directions respectively. The crystal 
dimensions along the $x$ and $z$ Cartesian directions are 10.76 nm and 3.85 nm respectively, 
whereas the $y$ dimension representing the film thickness is varied to study the surface 
and size effects. An edge dislocation is introduced at the centre of the film with dislocation 
line along the $z$ axis and Burgers vector $a$$<$111$>$/2 along the $x$ direction where the lattice constant 
$a$=0.31472 nm. Periodic boundary conditions are imposed on all the three directions, 
however, the boundaries are sufficiently extended along the $y$ direction so that free top and 
bottom surfaces can be created and interactions among the periodic image films can be eliminated. 
The dislocation core is identified by specifying a centro-symmetric deviation parameter 
window \cite{caibook} of width 0.024-0.1 nm$^2$. The system is then initialized at 300 K temperature. 
Precisely calculated forces are applied to the atoms of the top and bottom surfaces of the film with
directions parallel and antiparallel to the Burgers vector respectively so that a shear stress of 
250 MPa can be produced. Following a time lag, this applied stress is transmitted to the 
dislocation line, which in turn
attains a terminal velocity in several femtoseconds \cite{caibook} following Eqn. (1).
Constant temperature is maintained by implementing the Nos\'{e}-Hoover thermostat \cite{nose,hoover}. 
Trajectories of all the atoms are calculated at a time step of 0.5 fs. Positions of the 
dislocation core are recorded with respect to time and thus the dislocation velocity is extracted. 
Simulations are performed in two ways, case I: by reducing the film thickness equally about 
the dislocation line and case II: by varying 
the position of the dislocation line at different depths beneath the top surface of a film 
of fixed thickness.

Figure 3(a) shows the variation in the velocity of edge dislocation as a function of film thickness. 
The MD simulations clearly exhibit a significant increase in this velocity when the film thickness 
is reduced from $\sim$70 nm to 8.5 nm. However, at higher thicknesses the velocity of dislocation 
attains a constant value of $\sim$728 m/sec. A significant rise of 46$\%$ in this velocity for 
the film of 8.5 nm thickness under the same loading conditions is noteworthy in this context. 
A rising trend in the dislocation velocity is also observed as the dislocation line is brought 
closer to the top free surface in a film of 35.2 nm thickness (refer Fig. 3(b)).

The results obtained from the MD simulations establish the effects of surface and size on the velocities
of dislocations in thin films. These results have been used as a tool to separate out the contribution 
of lattice resistance from the overall drag. Equation (5) enables us to express the 
net drag coefficient $B$ as the sum of the drag coefficients due to lattice resistance 
$B_s^{(lattice)}$ and the remaining part $B^\prime$ due to other drags. $B$ values are calculated using
the velocities of dislocation extracted from MD simulations for two widely different film
thicknesses. The ratio of the drag coefficients corresponding to the lattice resistance 
for these two film thicknesses is evaluated and then used to separate out the constant part 
$B^\prime$ from $B$. 
The net drag coefficient $B$ for any arbitrary film thickness is obtained by 
combining the calculated values of $B^\prime$ and $B_s^{(lattice)}$ so that the respective  
velocity of dislocation can be evaluated. In order to perform these calculations, the
position vectors of the atoms are determined by superposing the following elastic displacement fields of 
an edge dislocation \cite{hirth} on a perefect bcc crystal; 

\begin{eqnarray}
u_x(x,y)= \frac{b}{2\pi}~[tan^{-1}\frac{y}{x} +\frac{xy}{2(1-\nu)(x^2+y^2)}], 
\end{eqnarray}

\begin{eqnarray}
u_y (x,y)=-\frac{b}{2\pi}~[\frac{1-2\nu}{4(1-\nu)}\ln(x^2+y^2)\nonumber\\ 
+\frac{x^2-y^2}{4(1-\nu)(x^2+y^2)}],
\end{eqnarray}

\begin{eqnarray}
u_z(x,y)= 0, 
\end{eqnarray}

\noindent
where the value of the Poisson's ratio $\nu$ is 0.3 \cite{hurley}. The change in the displacement 
fields of atoms due to the incremental change $\delta x$ in the
dislocation line position can be found to decay rapidly with the distance from the dislocation 
line as compared to the displacement field itself. Hence, the interactions among the periodic 
image dislocations are not expected to affect the results significantly. Thus, the numerical computations 
have been done taking into account 200 atoms on both sides of the dislocation line in each 
row along the $x$ direction. Number of rows and the position
of the origin are varied according to the configurations in case I and case II respectively.
The results of the calculations are shown in Fig. 3(a) and (b) in the form of solid lines. 
The model is found to reproduce the trends observed in MD simulations.  

Dislocation velocity at applied shear stress $\tau$ and temperature $T$ is empirically given as 
$v\sim \tau^m {exp(-Q/kT)}$ where $m$ is the stress exponent \cite{review,cai}. 
However, with rise in stress and temperature, the phonon drag becomes the predominant  
mechanism governing the dynamics of dislocations \cite{chang}. Since this phonon drag 
primarily constitutes $B^\prime$, the size effect on the velocity 
of dislocation should diminish at 
higher temperatures and applied stresses. Figure 4 represents the velocities of the edge dislocation 
obtained from the MD simulations for three different film thicknesses at three different applied 
loads as a function of temperature. The size effect on this velocity is noticeable only 
at 250 MPa shear stress and disappears at higher stresses of 500 MPa and 1000 MPa for the entire 
range of temperature studied. 
The MD simulations performed at 250 MPa stress show a decrease in the velocity of dislocation with 
increasing temperature. The reduction in the dispersion of 
the velocity of dislocation with thicknesses of thin films at higher stresses and temperatures 
is supportive of the fact that 
lattice resistance is the key factor behind the observation of the effects under discussion.

This letter reports a pronounced change in the velocity of a dislocation due to the presence 
of a free surface in the proximity of the dislocation line in a finite nanoscale crystalline solid. 
This effect has been attributed to the altered lattice resistance in different system configurations. 
A model following an unconventional approach to the lattice resistance has been developed that 
serves as a tool for explaining these observations. The fundamental ideas as well as the proposed 
model have been verified through the MD simulations of an edge dislocation in bcc molybdenum. 
Similar studies are yet to be done for more complex types of dislocations in different crystal 
structures and this largely simplified model provides ample scope of necessary modifications 
to suite these special cases. Development of a generalized model, especially the one covering 
up to the bulk regime needs further intensive study where the ideas as presented here, 
can provide a fundamental framework for the understanding of the mechanism of the lattice 
resistance along with the associated dislocation dynamics.

The authors thank Dr. Wei Cai for technical suggestions regarding the use of the MD++ molecular dynamics
package.\\

\newpage

\begin{figure}
\centerline{\includegraphics*[width=6.5cm,angle=270]{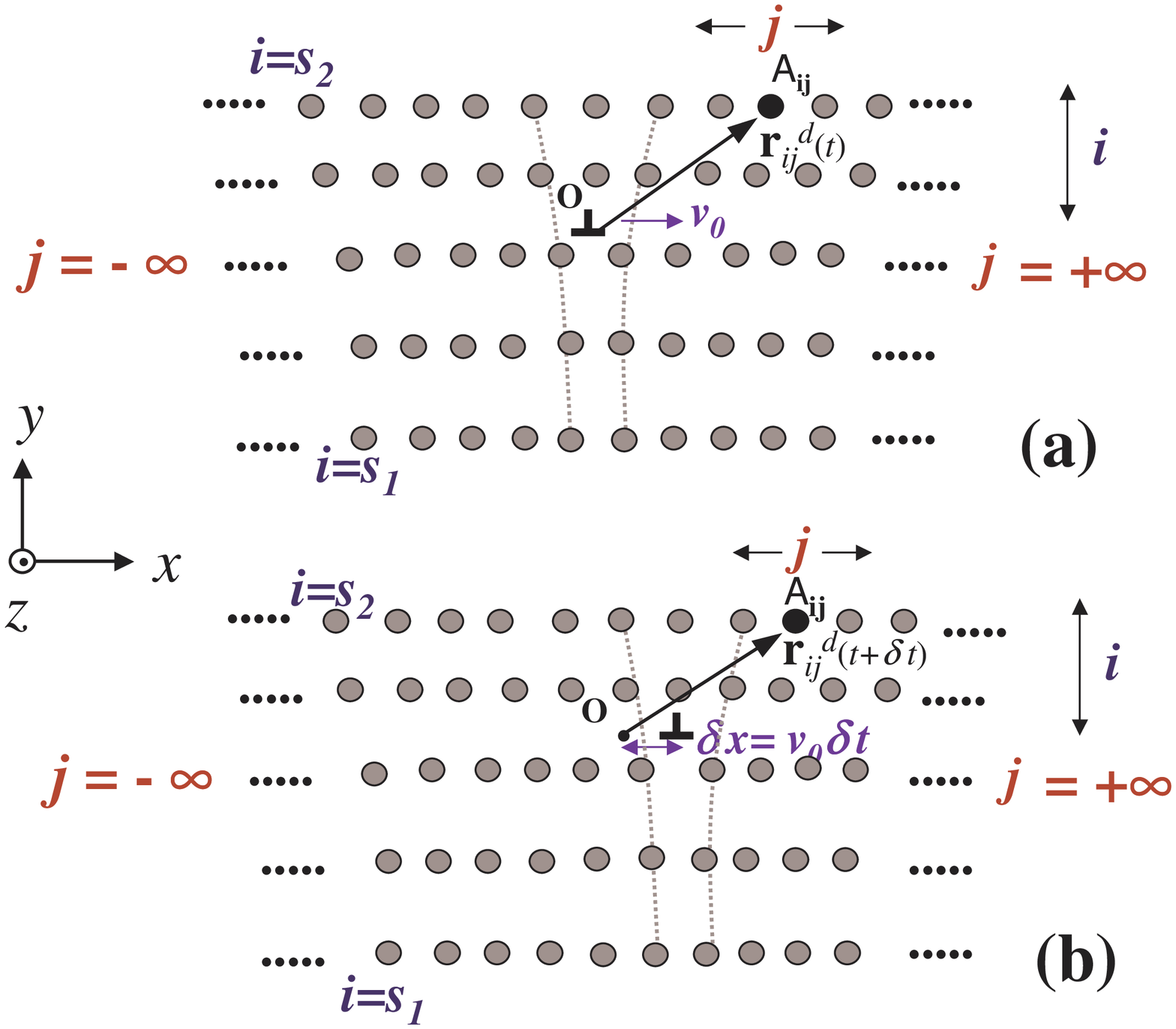}}
\caption{ Schematic representation of the model. 
(a) The edge dislocation line is along the $z$ axis and is at the origin O at time $t$.
A thin film is confined in $y$-dimension, where $i=s_1$ and $i=s_2$ are the free surfaces. 
The dislocation is moving with terminal velocity $v_0$ along the positive $x$ direction.  
${\bf r}_{ij}^d(t)$ is the position vector of the ${ij}^{th}$ atom $A_{ij}$. (b) In time $\delta t$,
the dislocation moves through $\delta x$ towards the positive $x$ direction and 
${\bf r}_{ij}^d (t+\delta t)$ is the position vector of the atom $A_{ij}$ at time $t+\delta t$.}
\end{figure}

\begin{figure}
\centerline{\includegraphics*[width=5.0cm,angle=270]{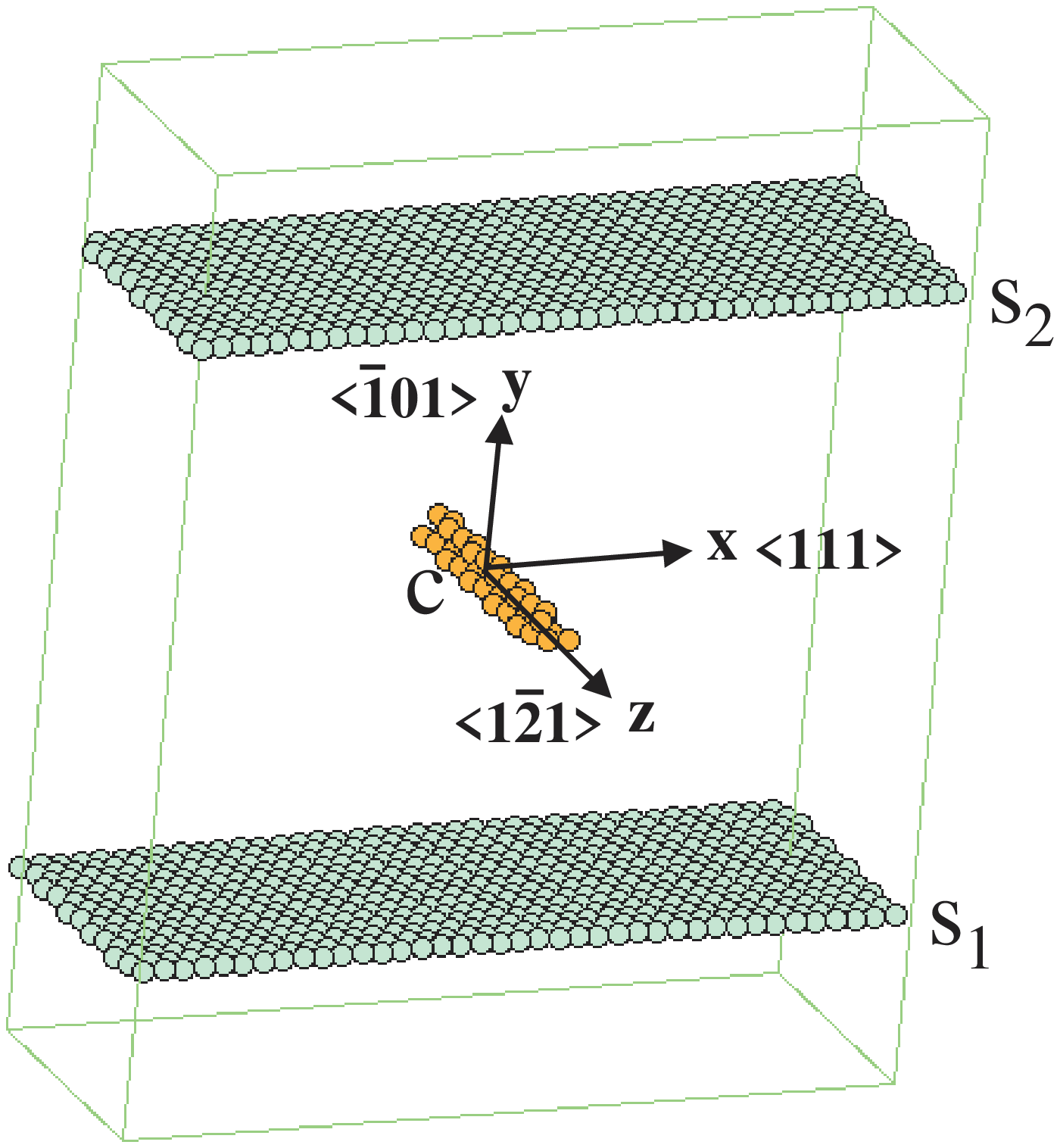}}
\caption{ The MD simulation cell with the indicated crystal directions. 
Only the atoms at surfaces ($s_1$, $s_2$) and the edge dislocation core (C) are displayed for clarity. 
Here the boundaries are extended along the $y$ direction.}
\end{figure}

\begin{figure}
\centerline{\includegraphics*[width=6.5cm,angle=270]{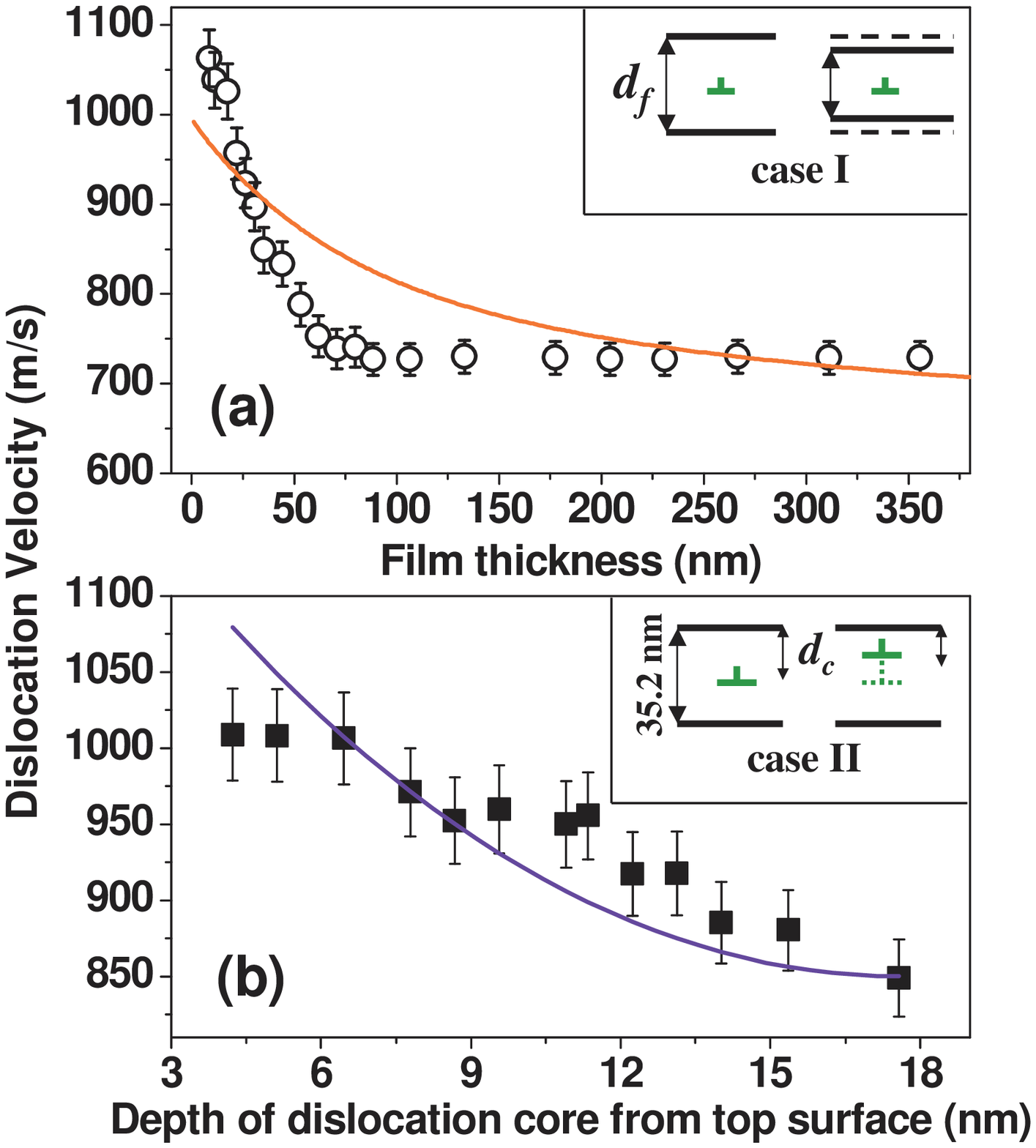}}
\caption{(color online). (a) The dislocation velocity obtained from MD simulations is 
plotted (circles) as a function of film thickness $d_f$ (case I). Here the dislocation 
line is equidistant from both the free surfaces of the film as illustrated schematically
in the inset. (b) The dislocation velocity extracted from MD simulations is presented (squares) 
for different depths of the dislocation line from the top surface ($d_c$)
of the film of 35.2 nm thickness (see case II in the inset). The solid lines represent the output 
of numerical calculations based on the model in both (a) and (b). Shear stress and temperature are 
250 MPa and 300 K respectively. 
Error bars for MD simulation results as indicated in both the plots are due to the randomness 
of the initial velocities and positions of the atoms in the simulation cell. }
\end{figure}

\begin{figure}
\centerline{\includegraphics*[width=4.5cm,angle=270]{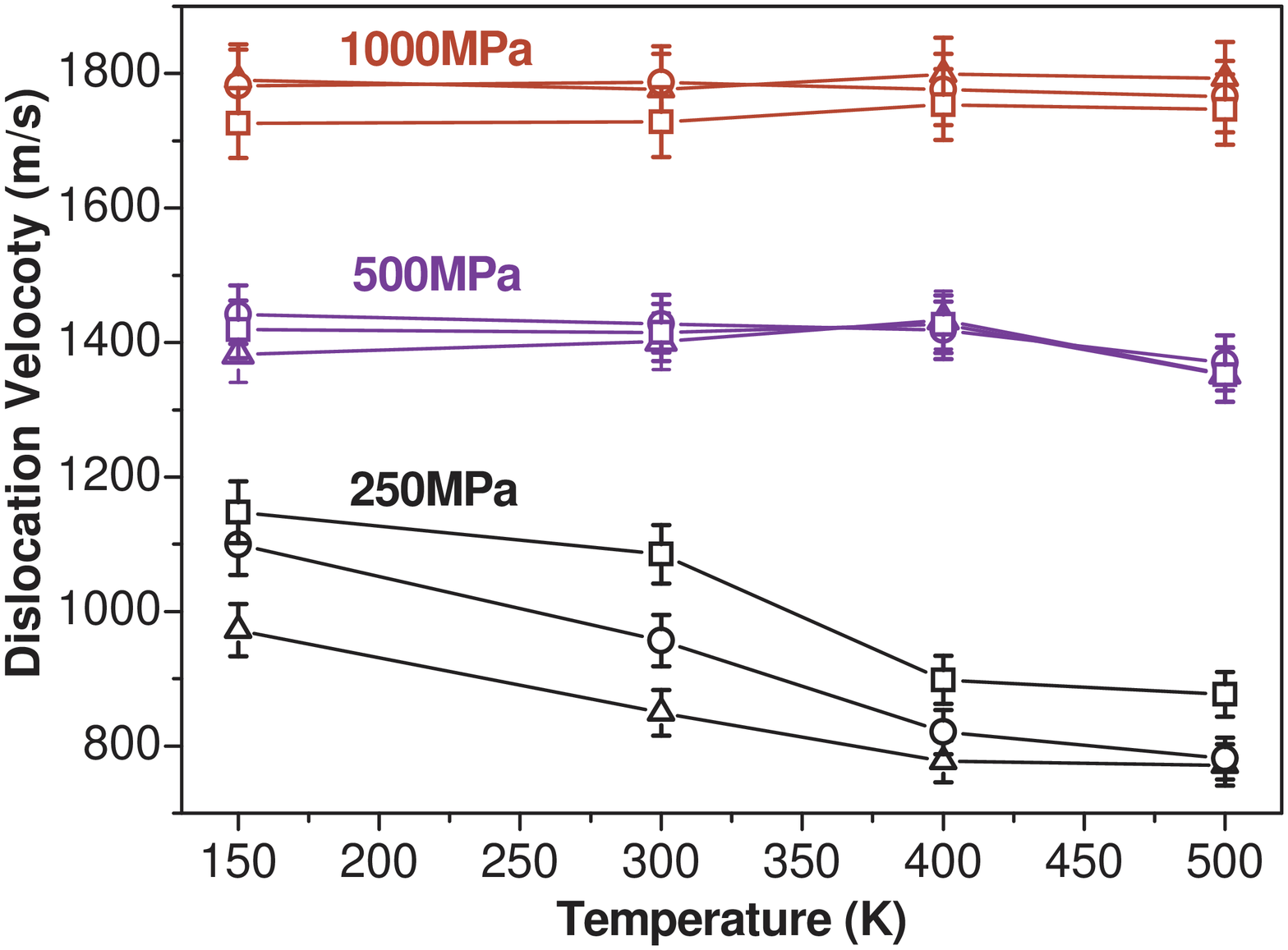}}
\caption{(color online). Variation of the dislocation velocities with temperature is plotted 
for three different film thicknesses 8.5 nm (square), 21.8 nm (circle) and 35.2 nm (triangle) 
at three different shear stresses of 250 MPa, 500 MPa and 1000 MPa. Error bars are indicated
in the figure. The dispersion of dislocation velocity with film thickness 
reduces at higher temperature and stresses.}
\end{figure}

\end{document}